\documentclass[journal=jacsat,manuscript=article]{achemso}

\usepackage[version=3]{mhchem} 
\usepackage{amssymb}

\author{Robert Stanton}
\email{stanton@lanl.gov}
\affiliation[Unknown University]
{Theoretical Division, Los Alamos National Laboratory, Los Alamos, NM 87545, USA}
\author{Mehmet Cagri Kaymak}
\affiliation[Unknown University]
{Theoretical Division, Los Alamos National Laboratory, Los Alamos, NM 87545, USA}
\author{Anders M. N. Niklasson}
\email{amn@lanl.gov}
\affiliation[Unknown University]
{Theoretical Division, Los Alamos National Laboratory, Los Alamos, NM 87545, USA}

\title[An \textsf{achemso} demo]
  {Shadow Molecular Dynamics for a Charge-Potential Equilibration Model}

\abbreviations{IR,NMR,UV}

\begin{document}







\begin{abstract}
      We introduce a shadow molecular dynamics (MD) approach based on the Atom-Condensed Kohn-Sham second-order (ACKS2) charge-potential equilibration model. In contrast to regular flexible charge models, the ACKS2 model includes both flexible atomic charges and potential fluctuation parameters that allow for physically correct charge fragmentation and scaling of the polarizability. Our shadow MD scheme is based on an approximation of the ACKS2's flexible charge-potential energy function, in combination with extended Lagrangian Born-Oppenheimer MD. Utilizing this shadow charge-potential equilibration approach mitigates the costly overhead and stability problems associated with finding well-converged iterative solutions to the charges and potential fluctuations of the ACKS2 model in an MD simulation. Our work provides a robust and versatile framework for efficient, high-fidelity MD simulations of diverse physical phenomena and applications. 
\end{abstract}

\section{1. Introduction}

    Molecular dynamics (MD) simulations are widely used in the computational investigation of material properties \cite{filipe2022molecular,yao2022applying,he2023advances,mahmood2022machine}. At its heart, an MD simulation involves calculating forces on individual atoms and integrating the equations of motion to capture the dynamics of the system. Naturally, the choice of the dynamical framework and the interatomic potential used to compute energies and forces must be carefully decided to faithfully represent the properties of interest. This leads to the central trade-off in MD simulations; balancing physical fidelity with computational cost \cite{sun2023probing}.
    Atomistic simulation frameworks can broadly be categorized in three classes, in order of decreasing computational complexity and fidelity: 1) \textit{ab initio} electronic structure approaches \cite{iftimie2005ab}, 2) semi-empirical quantum mechanical methods \cite{kulichenko2023semi,gerber2014ab}, and 3) techniques based on purely classical interatomic potentials \cite{brooks2021classical}. 
    Recently, machine learning (ML) methods have significantly advanced the ability to combine the low computational cost of purely classical approaches with the accuracy of quantum mechanical methods. However, this advancement often comes at the cost of reduced transferability \cite{mueller2020machine,zuo2020performance,montes2022training}. The limited transferability of classical ML-interatomic potentials (MLIPs) typically arises from their reliance on local descriptors and the fact that they incorporate minimal or no physical constraints to guide the dynamics beyond the training data\cite{kandy2023comparing,benoit2020measuring,wang2024machine,kulichenko2024data}. As a result, the  near \textit{ab initio} accuracy associated with MLIPs, is generally confined to systems closely related  to the training data,  where the dynamics evolve in well-sampled regions of phase space. Naturally, this leads to difficulties in simulating systems not well represented by the training data, such as those involving complex environments with long-range interactions, reactive processes, or dynamics in unexplored regions of the phase space. In such cases, the lack of robust physical constraints may become problematic \cite{ko2023recent,focassio2024performance}. 

    To address the transferability problem, while maintaining physical accuracy and a low computational cost, we need to develop efficient physics-informed models that allow for the rapid simulation of systems while satisfying imposed physical interactions and constraints.
    Physics-informed methods typically also offer a reduction in the number of model parameters 
    compared with purely deep-learning-based MLIPs \cite{pham2024neuromorphic,hu2023treating,zhou2022deep,karniadakis2021physics}. One set of physics-informed models, which is the focus of this article, are charge equilibration (QEq) models\cite{FJVesely77,Mortier1986,MSprik88,MSprik90,AKRappe91,DVanBelle92,WSRick94,TAHalgren01,GLamoureaux03,GAKaminsky04,PEMLopes09,PCieplak09,JZhifeng19,ongari2018evaluating,naserifar2017polarizable}. QEq models can be derived from a coarse-grained formulation of first-principles density functional theory \cite{DMYork96,GTabacchi02,ANiklasson21,JGoff23}. The interatomic potential from QEq models include an inexpensive charge-independent force field together with long-range Coulomb interactions between flexible, relaxed atomic charge densities. 
    QEq models provide a cheap alternative to fully orbital-resolved electronic structure calculations, while still offering predictive power and physical fidelity. The QEq family of methods is based on Sanderson’s principle of electronegativity equalization (EE) applied to atomistic simulations.\cite{Mortier1986}. Sanderson’s electronegativity equalization principles states simply that electrons in a molecule or crystal will flow such that the relative electronegativities throughout the system are brought to equilibrium. The QEq or EE models (used interchangeably) offer improved transferability with atomic partial charges approaching \textit{ab initio} accuracy to facilitate the simulation of large-scale complex molecular systems with drastically reduced computational costs compared to {\it ab initio} electronic structure calculations \cite{vondrak2023q,liu2022benchmarking,kadantsev2013fast}. 
    
    While the QEq approach provides an excellent foundation for the generation of high-fidelity, physics-informed models capable of simulating large extended systems, it still has some computational and physical drawbacks. 
    Computationally, solving a system of linear equations required by the QEq models can become prohibitively expensive for large systems, demanding the use of iterative Krylov-subspace solvers that must be tightly converged \cite{JGoff23}. Tight convergence is critical because approximate solutions can introduce non-conservative forces and numerical errors, leading to instabilities and systematic long-term energy drift \cite{niklasson2023shadow,kulichenko2023semi}. Physically, QEq models struggle to accurately describe integer charge fragmentations and tend to exhibit systematic deviations from the expected polarizability of dielectric systems \cite{acks2013}.  These shortcomings limit the reliability of QEq models in applications where high physical fidelity is required. In this work, we address both the computational and physical limitations of conventional QEq models by developing a shadow MD framework based on the 
    Atom-Condensed Kohn-Sham second-order (ACKS2) charge-potential fluctuation model
\cite{acks2013,verstraelen2014direct} The shadow MD approach improves the stability and reduces the computational cost, while the ACKS2 model extends the QEq framework by including not only flexible atomic partial charges but also potential fluctuation parameters, thereby enhancing the physical fidelity.
    
    Shadow potentials for MD simulations serve as a fairly general approach to design accurate, time reversible, and numerically stable methods that avoid the costly overhead of using iterative solvers. The key to the success of shadow MD-based approaches lies in replacing an exact potential (or energy function), for which only approximate quantities such as forces can be calculated, with an approximate shadow potential (or shadow energy function) that allows exact quantities to be computed directly and exactly, without relying on an iterative solver \cite{ANiklasson08, ANiklasson14,ANiklasson21b,niklasson2023shadow,henning2021shadowlagrangiandynamicssuperfluidity}.  The same underlying backward error analysis approach that this shadow MD is based on has been used in the past in the context of shadow Hamiltonian dynamics \cite{HYoshida90,CGrebogi90,SToxvaerd94,JGans00,BJLeimkuhler07,RDEngel05,SToxvaerd12}, from which we have borrowed the ``shadow'' terminology. However in this case, we use the terminology in the context of a dynamics driven by self-consistent non-linear models \cite{ANiklasson21b}. 
    
 Adopting the approximate shadow potential with its exact solutions to MD simulations reduces the computational cost and prevents error accumulation arising from non-conservative forces calculated by approximate iterative solvers.
 In order to maintain faithfulness to the exact potential, additional dynamical degrees of freedom (DOFs) are propagated in an extended Lagrangian (XL) approach that ensure solutions to the approximate shadow potential remain near those of the exact potential \cite{niklasson2009extended,niklasson2020extended,ANiklasson21b}.
 This allows for significantly improved numerical stability, and also facilitates time-reversibility, whereby the instabilities and energy drift associated with inexact forces can be avoided \cite{DRemler90,PPulay04,ANiklasson07}.
 An important aspect of the shadow potential approach is also the freedom of choice in the functional form that the shadow potential takes. In particular, we can design approximate shadow potentials for which efficient exact solutions exist at little cost. This can drastically improve the computational efficiency of the MD simulations \cite{ANiklasson21b,niklasson2020density}.

    The original formulation of the regular QEq approach has some limitations in its ability to capture the charge distribution, namely: 1) unphysical fractional charges in isolated molecular fragments; and 2) deviation in expected macroscopic polarizabilities in dielectric systems. Several approaches or extensions have been developed to mitigate these issues, as summarized by Verstraelen \textit{et.\ al} upon their introduction of the ACKS2 framework \cite{acks2013,verstraelen2014direct,patel2004charmm,lee2008origin,chelli2005polarization,chen2007qtpie,morales2001classical,chelli1999electrical,halgren1996merck,nistor2006generalization}. The ACKS2 framework extends the QEq formalism by including not only a set of flexible atomic charges, but also on-site potential fluctuations, which modulate the ease of charge transfer between atoms.  In this work, we introduce a shadow MD approach based on the ACKS2 model. To clearly separate the shadow MD from the original ACKS2 model in our presentation, we will refer to our modified approach as the shadow charge-potential equilibration (SChPEq) framework. This  SChPEq framework for MD simulations provides excellent stability and computational efficiency, while also maintaining close agreement with the underlying `exact' ACKS2 reference potential.



\section{2. The Shadow Charge-Potential Equilibration Framework for MD Simulations}

To overcome the computational and physical shortcomings of Born-Oppenheimer MD simulations based on regular QEq models, we propose a shadow MD approach using a charge-potential equilibration model based on the ACKS2 framework. For clarity, throughout the manuscript we refer to regular QEq models as the family of frameworks where the charge-dependent energy is described by a second-order expansion in the net partial charges,
\begin{equation}
    E({\bf q}) = {\boldsymbol{\chi}}^{\rm T} {\bf q} + \frac{1}{2}{\bf q}^{\rm T}{\bf C}{\bf q},
\end{equation}
using fixed electronegativities $\bf \chi$ and chemical hardness parameters for each atom type.
Here $\bf q$ represents the on-site partial charges, $\bf C$ is the Coulomb operator, where its diagonal on-site elements correspond to the hardness parameters. The equilibrated ground state solution, under the condition of a net partial charge, $\sum_i q_i = Q_{\rm tot}$, is then given by the solution to a system of linear equations,

\begin{equation}
~~\begin{bmatrix}
{\bf C}& {\bf 1}\\
{\bf 1^T} & {0} \\
\end{bmatrix}
\begin{bmatrix}
{\bf q}\\
{\bf \mu}\\
\end{bmatrix}
=
\begin{bmatrix}
-{\boldsymbol \chi}\\
Q_{\rm tot}\\
\end{bmatrix}.
\label{eq:qeq}
\end{equation}
Here $\mu$ is a Lagrange multiplier enforcing the correct net charge of the system.

In this section we introduce the original ACKS2 framework and the new SChPEq model together with the corresponding extended Lagrangian Born-Oppenheimer MD. These models extend the above-defined QEq framework to include not only atomic partial charges, but also fluctuations in the on-site potential of each atom. 

\subsection{2.1 The ACKS2 Framework}

The ACKS2 model is based on a second-order expansion of the electronic energy, $E({\bf q}, {\bf v})$, in flexible atomic net partial charges, ${\bf q} \equiv \{q_i\}$, and potential fluctuations, ${\bf v} \equiv \{v_i\}$. 
In a matrix-vector notation the ACKS2 electronic energy can be expressed as
\begin{equation}
    E({\bf q}, {\bf v}) = {\boldsymbol \chi}^{\rm T} {\bf q}  + ({\bf q}-{\bf q}_{\rm 0})^{\rm T} {\bf v} + \frac{1}{2}{\bf q}^{\rm T} {\bf C}{\bf q} + \frac{1}{2}{\bf v}^T{\bf X}{\bf v}, \label{ACKS2Energy}
\end{equation}
where ${\boldsymbol \chi} \equiv \{\chi_i\}$ are the atomic electronegativities and ${\bf C}$ and ${\bf X}$ (expanded on later in this section) are the Coulomb and potential fluctuation interaction matrices, respectively. 
The Born-Oppenheimer potential is then given by the constrained optimization,
\begin{equation}
    U_{\rm BO}({\bf R}) = V_{\rm S}({\bf R}) + \operatorname{stat}_{{\bf q},\bf v} \Big \{ E({\bf q}, {\bf v}) ~\Big \vert \sum_i v_i = V_{\rm tot}, ~ \sum_i q_i = Q_{\rm tot} \Big\}. \label{eq:BO_min}
\end{equation}
Here $V_{\rm S}({\bf R})$ is a short-range, charge-independent potential as a function of the atomic coordinates, ${\bf R} = \{{\bf R}_i\}$, and ${\rm stat}_{\bf q,v}$ denotes the stationary solution ($\min$ and $\max$) with respect to variations in ${\bf q}$ and ${\bf v}$. The potential fluctuation parameters, ${\bf v}$, act as a set of Lagrange multipliers that constrain the charges to the chosen reference charges ${\bf q}_{\rm 0}$ when ${|\bf X|} \rightarrow {\bf 0}$, and when an individual atom $i$ becomes spatially separated from all other atoms in the system its charge becomes ${{q}_0}_i$. In this way it is easy to control the fragmentation limit both for the full system as well as for the separation of individual atoms.
From the Born-Oppenheimer potential energy surface we get the dynamics from Newton's equation of motion,
\begin{equation}
    M_i {\bf \ddot R}_i = - \nabla_i U_{\rm BO}({\bf R}), \label{eq:Newton}
\end{equation}
where $M_i$ is the atomic mass at position ${\bf R}_i$.

The solution to the constrained ACKS2 optimization problem in Eq.\ (\ref{eq:BO_min}) is given by a system of linear equations,
\begin{equation}
~~\begin{bmatrix}
{\bf C}& {\bf I} & {\bf 1} & {\bf 0}\\
{\bf I} & {\bf X} & {\bf 0} & {\bf 1}\\
{\bf 1}^{\rm T} & {\bf 0} & { 0} & { 0}\\
{\bf 0} & {\bf 1}^{\rm T} & { 0} & { 0}
\end{bmatrix}
\begin{bmatrix}
{\bf q}\\
{\bf v}\\
\mu\\
\lambda
\end{bmatrix}
=
\begin{bmatrix}
-{\boldsymbol \chi}\\
{\bf q}_{\rm 0}\\
Q_{\rm tot}\\
V_{\rm tot}
\end{bmatrix}.
\label{eq:1}
\end{equation}
    Here ${\bf I}$ denotes the identity matrix of shape $N\times N$, ${\bf 1}$ denotes the column vector of $1$'s of length $N$. The variables
    $\mu$ and $\lambda$ are the corresponding Lagrange multipliers that ensure the conservation of the sum of $v_i$ and $q_i$, respectively. 
    The charges ${\bf q}_{\rm 0}$ determine the (typically integer) net charges to be approached asymptotically for isolated atoms. The Coulomb interaction matrix $\bf C$ is given by $C_{ij}=\delta_{ij}\eta_i + (1-\delta_{ij})\Gamma_{ij}$, where $\Gamma_{ij}$ 
    has the form $\Gamma_{ij}=\left( r_{ij}^3 + \gamma_{ij}^{-3}\right)^{-\frac{1}{3}}$ with $r_{ij} = |{\bf r}_i - {\bf r}_j|$. Here $\eta_i$ represents the on-site chemical hardness, and $\gamma_{ij}$ is the geometric mean of an atom-wise defined shielding parameter for the Coulomb interaction at short distances. The ${\bf X}$ matrix captures the interactions between the potential fluctuations, and serves to facilitate or suppress the exchange of charge between atoms. This term appears like a chemical softness that mitigates some shortcomings of the standard QEq approach such as fractional charges for isolated chemical fragments and unphysical scaling of the polarizability. Off-diagonal elements of the ${\bf X}$ are nonzero only within a cutoff radius $r_{ij}<r_{\rm cut}$ and take the form; $X_{ij}=\Lambda \left( \frac{r_{ij}}{\sigma_{ij}} \right)^3 \left( 1- \frac{r_{ij}}{\sigma_{ij}} \right)^6$. Diagonal elements of ${\bf X}$ are the negative row-sum of the off-diagonal elements, $\sigma_{ij}$ is the average of the pairwise parameters setting the range of the bond softness, and $\Lambda$ is a global bond softness parameter. This choice of approximating the ${\bf X}$ matrix follows from the implementation of the ACKS2 framework to the study of Li$_2$O slabs, as detailed by O'Hearn \textit{et. al} \cite{o2020optimization}. In this work, obtaining a sufficiently applicable parameterization represents a large portion of the research efforts, illustrating the need for automized parameterization pipelines. Apart form the studies of Li$_2$O slabs, the ACKS2 model has also been used, for example, to simulate the co-pyrolysis of coal and polystyerene, and even the explicit charge transfer dynamics of free electrons in both molecular organic radicals and solid oxides \cite{wu2022insight,hossain2024ereaxff}. Additionally, it has been used to model charge response kernels and polarizabilities \cite{Shao_2022, bronson2024improving}, among other various applications \cite{koski2021water}.

The ACKS2 system provides a substantial improvement in computing atomic partial charges and polarizations, particularly in systems containing isolated molecular fragments. Notably, in the case of separate fragments ACKS2 outperforms non-spin polarized first-principles DFT and semi-empirical quantum mechanical approaches, while delivering a significant increase in computational efficiency. Compared to QEq methods, the ACKS2 model requires the inversion of a matrix of twice the size, i.e.\ see Eqs.\ (\ref{eq:qeq}) and (\ref{eq:1}). When pairing ACKS2 with charge neutral force fields such as ReaxFF or MLIPs, the solution of this system of linear equations can rapidly become the computational bottleneck of the approach. Furthermore, the utilization of iterative solvers with insufficient convergence introduces approximate non-conservative forces and numerical errors that accumulate over time, leading to instabilities that may invalidate the results of the MD simulations. This is a particular problem when an initial guess that is extrapolated from previous time steps is used to accelerate the iterative solver. Several techniques based on extended Lagrangian Car-Parrinello MD \cite{RCar85,MSprik88,MSprik90} or more recently extended Lagrangian Born-Oppenheimer MD (XL-BOMD) \cite{ANiklasson08,ANiklasson21b}, which originally was developed for first-principles quantum-based MD have been developed to mitigate these shortcomings for polarizable force field models, e.g.\ see \cite{KNomura15,AAlbaugh15,AAlbaugh17,AAlbaugh18,ILeven_19,DAn20}. Here we will follow the idea behind XL-BOMD but in combination with an approximate shadow potential formulation for the ACKS2 model \cite{ANiklasson21,ANiklasson21b,JGoff23,CHLi2025}. This technique is based on more recent developments of XL-BOMD \cite{niklasson2020extended} and avoids any need for an iterative solver. 
    We denote this approach the shadow charge-potential equilibration (SChPEq) model, which maintains excellent agreement with ACKS2 while offering reduced improved computational cost and long-term stability in MD simulations.

\subsection{2.2 The SChPEq Framework}

     The SChPEq framework is based on an approximation of the charge-dependent ACKS2 energy function. The energy for this approximate 'shadow' energy function at the equilibrated minima determines our approximate 'shadow' Born-Oppenheimer potential. There are many options to construct an approximate shadow energy function. However, certain conditions need to be fulfilled such that a unique solution for the relaxed charges exist that closely follows the exact fully optimized state for the regular ACKS2 problem. It is also important that the exact ground state solution for the relaxed charges that determines the shadow potential is easy to calculate without a costly iterative procedure.

     Our approximate shadow energy function, ${\cal E}({\bf q,v},{\bf n,u})$, is given by a partial linearization of the ACKS2 energy expression in Eq.\ (\ref{ACKS2Energy}) in ${\bf q}$ and ${\bf v}$ around an approximate solution, ${\bf n}$ and ${\bf u}$ to the exact optimized ACKS2 charges from Eq.\ (\ref{eq:1}). i
     The shadow energy function is defined by
     \begin{equation}
{\cal E}({\bf q,v},{\bf n,u}) = 
{\boldsymbol \chi}^{\rm T}{\bf q}  - {\bf q}_{\rm 0}^{\rm T}{\bf v} + \frac{1}{2} [{\bf q}^{\rm T}{\bf v}^{\rm T}] 
\begin{bmatrix}
{\bf C}_{\rm S}& {\bf I}\\
{\bf I} & {\bf X}_{\rm S}
\end{bmatrix}
\begin{bmatrix}
{\bf q}\\
{\bf v}
\end{bmatrix} + \frac{1}{2} \left(2[{\bf q}^{\rm T}{\bf v}^{\rm T}] - [{\bf n}^{\rm T}{\bf u}^{\rm T}] \right)
\begin{bmatrix}
{\bf C}_{\rm L}& {\bf 0}\\
{\bf 0} & {\boldsymbol X}_{\rm L}
\end{bmatrix}
\begin{bmatrix}
{\bf n}\\
{\bf u}
\end{bmatrix}
\label{eq:3}
\end{equation}
Here we have split the Coulomb operator in a short-range (S) (on-site, diagonal) and a long-range (L) (non-local, off-diagonal) part, where $\bf{C} = \bf{C}_{\rm S} + \bf{C}_{\rm L}$. The same split is performed for the potential fluctuation matrix, where $\bf{X}=  \bf{X}_{\rm S} + \bf{X}_{\rm L}$. The separation of the Coulomb and potential fluctuation matrices into an on-site and non-local part allows for the partial linearization of the system of equations for the off-diagonal components, whereas the diagonal on-site terms are kept to second-order. The charge-potential fluctuation optimization problem that defines the corresponding shadow Born-Oppenheimer potential of our SChPEq model, is given by,
\begin{equation}
    {\cal U_{\rm BO}}({\bf R,n,u}) = V_{\rm S}({\bf R}) + {\rm stat}_{{\bf q},{\bf v}} \left\{ {\cal E}({\bf q,v},{\bf n,u})  \Big \vert  \sum_i q_i =Q_{\rm tot}, \sum_i v_i =  V_{\rm tot} \right\}.
\label{eq:4}
\end{equation} 
The solution to this constrained optimization problem is determined by a  quasi-diagonal set of linear equations,
\begin{equation}
~~\begin{bmatrix}
{\bf C_{\rm S}}& {\bf I} & {\bf 1} & {\bf 0}\\
{\bf I} & {\bf X_{\rm S}} & {\bf 0} & {\bf 1}\\
{\bf 1}^{\rm T} & {\bf 0} & { 0} & { 0}\\
{\bf 0} & {\bf 1}^{\rm T} & { 0} & { 0}
\end{bmatrix}
\begin{bmatrix}
{\bf q}({\bf n,u})\\
{\bf v}({\bf n,u})\\
\mu\\
\lambda
\end{bmatrix}
=
\begin{bmatrix}
-{\boldsymbol \chi} - \bf C_{\rm L} n\\
{\bf q}_{\rm 0}- \bf X_{\rm L} u\\
Q_{\rm tot}\\
V_{\rm tot}
\end{bmatrix}.
\label{eq:schpeq}
\end{equation}
If ${\bf C}_{\rm S}$ and ${\bf X}_{\rm S}$ are both diagonal this set of equations can be solved exactly with a direct method at very little cost, without relying on an iterative solver. This can be achieved by using the Woodbury formula, where the rows and columns with ${\bf 1}$s correspond to a rank-4 update and where the the diagonal blocks of ${\bf C}_{\rm S}$ and ${\bf X}_{\rm S}$ together with the off-diagonal ${\bf I}$ blocks corresponds to a set of independent $2\times 2$ matrices. In this way Eq.\ (\ref{eq:schpeq}) can be solved directly at little cost compared to the ACKS2 problem in Eq.\ (\ref{eq:1}).

The exact solution to the shadow problem in our SChPEq model will deviate from the solution to original regular ACKS2 set of equations in Eq.\ (\ref{eq:1}). 
The accuracy of the SChPEq potential, ${\cal U_{\rm BO}}({\bf R,n,u})$, for some configuration, ${\bf R}$, will depend on how close ${\bf n}$ and ${\bf u}$ are to the exact solution, i.e.\ ${\bf q}$ and ${\bf v}$, of the original ACKS2 problem in Eq.\ (\ref{eq:1}) for that same geometry. If we have chosen ${\bf n}$ and ${\bf u}$ to be the exact solution at a given geometry, ${\bf R}^{(0)}$, the SChPEq potential, ${\cal U_{\rm BO}}({\bf R,n,u})$ will be accurate only in the vicinity of that configuration. This can be understood from the fact that ${\bf n}$ and ${\bf u}$ are the points around which we expand the ACKS2 energy function to construct the shadow energy function ${\cal E}({\bf q,v},{\bf n,u})$, which defines our SChPEq potential in Eq.\ (\ref{eq:4}).\\

\begin{figure}[h]
    \caption{The SChPEq potential energy surface linearized about the exact solution at the equilibrium geometry of a water molecule  in comparison to the corresponding ACKS2 potential (top panel a) as a function of the bond length of one of the water molecules. The lower panel (b) shows a SChPEq potential that was obtained using a linearization about a stretched configuration.  Only the charge-dependent part, $V_{\rm charged}$ of the potential energies are shown.}
\centering
\includegraphics[width=1.0\textwidth]{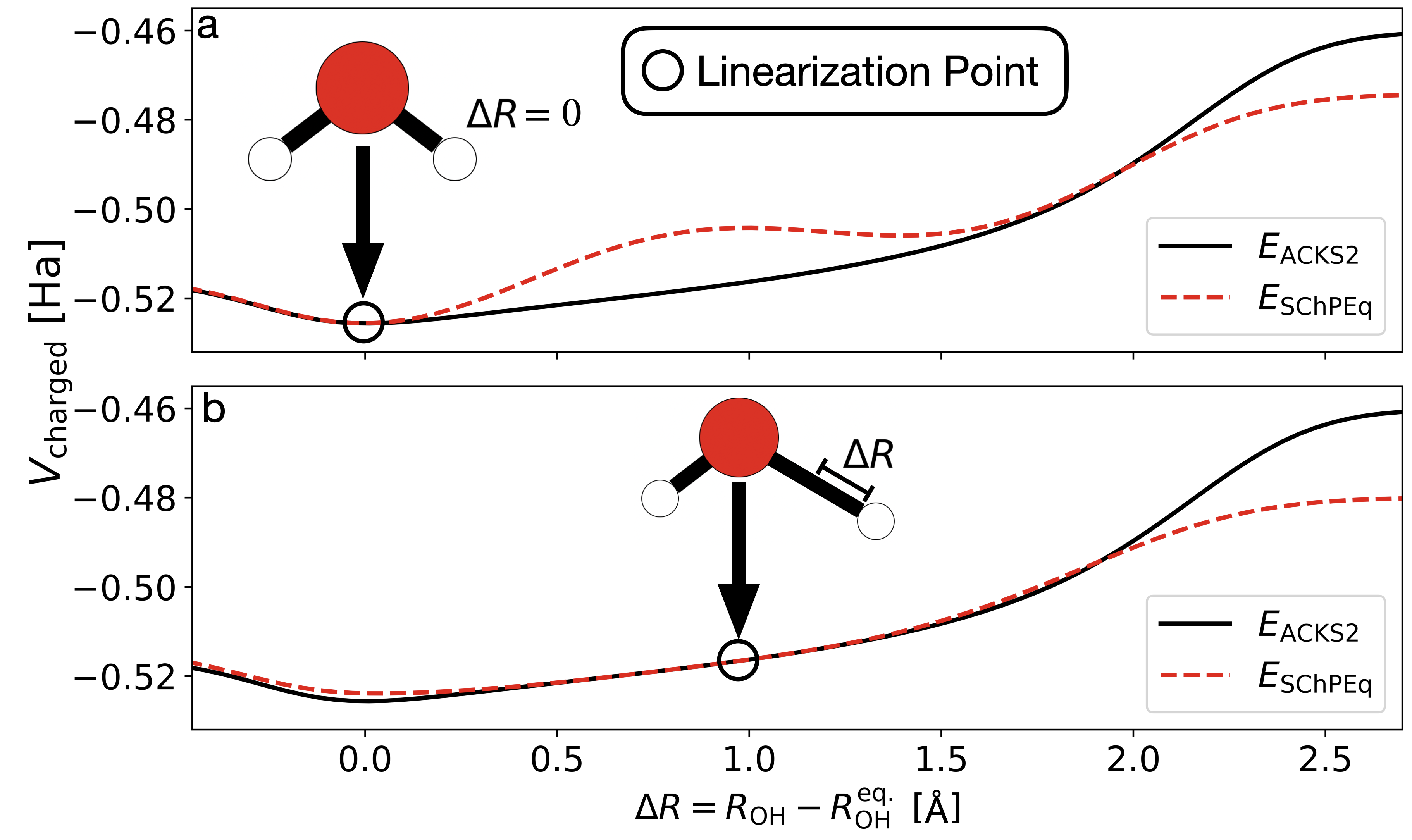}
\label{fig:linearize}
\end{figure}

An important conceptual note is that each choice of where we chose to carry out the linearization around the exact solution generates an entirely new shadow potential energy surface. To illustrate this, we consider a scan along the O-H bond coordinate of a water molecule where the shadow energy function is linearized either around the exact charges and potential parameters at the equilibrium geometry or around a stretched configuration. 
In Fig. \ref{fig:linearize}a (upper panel), the charge-dependent portion of the SChPEq potential based on a linearization around the exact solution at the equilibrium O-H bond length is shown for a water molecule, as well as the corresponding charge-dependent exact ACKS2 reference potential. The bottom panel (Fig. \ref{fig:linearize}b) demonstrates the SChPEq potential upon linearization around a stretched O-H bond.
The key insight from this analysis is that the SChPEq potentials can differ largely depending on where the linearization is carried out, emphasizing the fact that the shadow potential is only valid in the vicinity of where $\bf n$ and $\bf u$ accurately mirror $\bf q$ and $\bf u$ of the exact potential.
The SChPEq and ACKS2 potentials are thus nearly identical about the linearization point, but may deviate significantly as one moves away.
While this is difficult to show analytically, the subsequent sections show \textit{a posteriori} that this conclusion holds throughout all systems tested in this work.
In other words, these results show that the highest accuracy of the SChPEq potential energy surface is only given in the region around which the linearization has occured. As the system evolves in a MD simulation we therefore need to ensure that $\bf n$ and $\bf u$ of the SChPEq potential somehow remain close to ${\bf q}$ and ${\bf v}$ of the underlying exact ACKS2 potential for each new geometry.

To keep ${\bf n}$ and ${\bf u}$ close to the exact ACKS2 solution we can incorporate them as additional dynamical variables, ${\bf n}(t)$ and ${\bf u}(t)$, that are propagated alongside the atomic positions, ${\bf R}(t)$, and velocities, ${\bf \dot R}(t)$. We do this by using a harmonic oscillator that is centered around a close approximation to the exact ACKS2 solution. We can formulate the dynamics of our SChPEq model with an extended Lagrangian that we define as
\begin{equation}
\begin{aligned}
    {\cal L}({\bf R, \dot{R}, n, \dot{n}, u, \dot{u}}) = \frac{1}{2} \sum_i M_i|{\bf{\dot{R}}}_i|^2 - {\cal U_{\rm BO}}({\bf R, n, u}) + \frac{1}{2} m \sum_i (\dot{n_i}^2 + \dot{u_i}^2) \\
    - \frac{1}{2}m \omega^2\left( [{\bf q}^{\rm T}({\bf n,u})~ {\bf v}^{\rm T}({\bf n,u})] - [{\bf n}^{\rm T}~{\bf u}^{\rm T}] \right){\bf T}  
    \begin{bmatrix}
     {\bf q}({\bf n,u}) - {\bf n}\\ 
     {\bf v}({\bf n,u}) - {\bf u}\\ 
     \end{bmatrix} .
\end{aligned}
\label{eq:XL}
\end{equation}
Here $m$ is a fictitious mass parameter for the electronic and potential fluctuation degrees of freedom, $\omega$ is the frequency of the harmonic oscillator in the last two terms, and ${\bf T} = {\bf K}^{\rm T}{\bf K}$  is a symmetric positive metric tensor given as the square of a kernel, ${\bf K}$. We define this kernel as the inverse of the Jacobian, ${\bf J}$, of the residual functions in the harmonic potential. Using the composite vector notation, where ${\bf c}[{\bf x}] \equiv [{\bf q}({\bf n,u}) ~{\bf v}({\bf n,u})]$ with ${\bf x} \equiv [{\bf n ~u}]$, the residual function takes the form ${\bf f}({\bf x}) = {\bf c}[{\bf x}] - {\bf x}$. The Jacobian is then given by 
\begin{equation}
    J_{ij} = \frac{\partial f_i({\bf x})}{\partial x_j} \equiv \frac{\partial (c_i[{\bf x}]-x_i)}{\partial x_j}.
\end{equation}
The exact full Jacobian would be expensive to calculate explicitly. However, a preconditioned low-rank approximation can be used \cite{niklasson2020extended,niklasson2020density,CHLi2025}, which keeps the computational cost low, as will be described further below. 

\subsection{2.3 Extended Lagrangian Dynamics in the SChPEq Framework}

The extended Lagrangian in Eq.\ (\ref{eq:XL}) generates equations of motion given by the Euler-Lagrange equations for the extended Lagrangian, ${\cal L}$. We derive these equations in an adiabatic limit, were we assume that the frequency, $\omega$, of the extended harmonic oscillator is high compared to the fastest nuclear motion and that the mass of the electronic degrees of freedom, $m$, is small. We can take this classical adiabatic, Born-Oppenheimer-like limit, when we derive the equations of motion from ${\cal L}$, by letting $\omega \rightarrow \infty$, $\mu \rightarrow 0$, as $\mu \omega = {\rm constant}$ \cite{ANiklasson21b}. In this adiabatic limit we find the equation of motion that defines our SChPEq model, where
\begin{align}
    & M_i {\bf \ddot R}_i = - \frac{\partial {\cal U}_{\rm BO}({\bf R,n,u})}{\partial {\bf R}_i}\Big \vert_{{\bf n,u}},
\end{align}
and
\begin{align}
    & \begin{bmatrix}
        {\bf \ddot n}\\
        {\bf \ddot u}
    \end{bmatrix} = - \omega^2 {\bf K}\begin{bmatrix}
        {\bf q}({\bf n,u})- {\bf n}\\
        {\bf v}({\bf n,u})- {\bf u}
    \end{bmatrix}.
\end{align}
These two equations together with the calculation of the shadow potential in Eqs.\ (\ref{eq:4}) and (\ref{eq:schpeq})
are the key results of this paper that define our SChPEq MD simulation framework for charge-potential fluctuation models.
The first equation of motion for the atomic positions is very similar to the original Newton's equation of motion in Eq.\ (\ref{eq:Newton}). However, the ACKS2 Born-Oppenheimer potential is replaced by the approximate shadow potential, which is constructed through an exact direct optimization of the approximate shadow energy function, Eq.\ (\ref{eq:3}). For this shadow potential it is easy to calculate the correct conservative forces from the gradient of the shadow potential. The gradient is taken under constant ${\bf n}$ and ${\bf u}$, because they are included as dynamical variables in the Lagrangian formalism.
In this way the problem with finding a sufficiently converged and costly iterative solution that determines the regular ACKS2 Born-Oppenheimer potential and the corresponding forces is avoided. This greatly reduces the computational overhead and improves long-term stability of the SChPEq MD simulations.

The second set of equations of motion of the SChPEq model are harmonic oscillators for the extended charge and potential fluctuation degrees of freedom. Their purpose is to propagate ${\bf n}(t)$ and ${\bf u}(t)$ such that they provide a reasonably good approximation to the exact ground state solution of the ACKS2 Born-Oppenheimer problem. This allows plenty of flexibility in the integration of the equations of motion. In particular, the kernel, ${\bf K}$, can be approximated using a preconditioned low-rank Krylov subspace expansion \cite{niklasson2020extended,ANiklasson21b} and the convergence of this expansion does not need to be very tight.

To integrate the equations of motion we use a leapfrog velocity Verlet scheme for the nuclear degrees of freedom and a modified Verlet scheme for the integration of the extended charge and potential fluctuation degrees of freedom \cite{niklasson2009extended,PSteneteg10,GZheng11}. The modified Verlet integration scheme includes a weak dissipative force term (given by a linear combination of the current and six previous time steps) that keeps the electronic degrees of freedom synchronized with the motion of the atomic positions and removes the accumulation of numerical noise.

\section{3 Results and Analysis}

To evaluate the quality of the SChPEq model we need to compare it to the `exact' ACKS2 reference data during MD simulations. The purpose with this analysis is not to accurately reproduce first-principles calculations and experimental data or to demonstrate the best possible computational performance, but to show that the underlying ACKS2 model is accurately reproduced with the proposed shadow MD. 

In our MD simulations, the charge neutral part of the total energy, $V_{\rm S}({\bf R})$, is computed using a charge-independent tight-binding energy term from the GFN0-xTB implementation in the xTB quantum chemistry code with the electrostatic energy, $E_{\rm IES}$, turned off \cite {CBannwarth19,bannwarth2021extended,grimme2017robust}.

This term is then added to the non-local charge dependent part of the potential energy given from the contrained equilibration of the corresponding charge or charge-potential energy function.     The charge-potential equilibration scheme presented here was developed in-house using Python and integrated with xTB and the Atomic Simulation Environment using the xTB-ASE interface \cite{larsen2017atomic}. Additionally, we note that for the propagation of the XL-DOFs, we have employed a 3$^{\rm rd}$ rank update for kernel $\bf K$ after using a constant preconditioner, $\bf K_0$, calculated only once at the initial, first time step.

The values of the parameters used in QEq models, semi-empirical QM approaches, or machine learned force fields for materials science are typically determined as an optimization problem, which aims to minimize the residual between the values of some chosen set of properties and those obtained from some reference method or experiments. Common quantities used in this optimization (or training) are total energies, atomization energies, forces, atomic partial charges, or molecular dipoles \cite{deng2023chgnet,luo2024mepo,li2025shadow,zhou2022deep}. 
All learned parameters of our charge-potential equilibration model were obtained through particle swarm optimization with respect to atomic partial charges calculated form first-principles density functional theory (DFT) \cite{hohen,RParr89,RMDreizler90}(see the Computational Methods for more additional details), with all particles sampled from a uniform distribution over parameter ranges which lead to reasonable atomic charges. In the case of the water systems, DFT-based MD simulations were carried out at 300 K with the Andersen thermostat, and atomic charges were computed along this trajectory to obtain 50 uncorrelated molecular configurations. The formulation of more robust and automized parameterization pipelines may form the basis for future work to further improve the accuracy and analysis of more advanced and complex, large-scale systems.

\subsection{3.1 Accuracy of the SChPEq model}

By integrating the extended Lagrangian degrees of freedom (XL-DOFs), i.e.\ ${\bf n}(t)$ and ${\bf u}(t)$, using different sizes of the integration time step, $\delta t$, we can analyze the accuracy and scaling of the error compared to the 'exact' ACKS2 model in the SChPEq MD simulations. We make this comparison for the Born-Oppenheimer potential energy, the forces, the charges, and the potential fluctuations.
The theoretically expected scaling as a function of $\delta t$ for the error in the SChPEq potential is $\Delta E \propto \delta t^4$, and for the forces, $\delta F \propto \delta t^2$ \cite{ANiklasson21b}. The error in the charges and potential fluctuation are expected to scale as $\delta t^2$.

\begin{figure}[h]
    \caption{The size and scaling of the difference in the a) potential energy, b) force, c) charge, d) and potential as a function of the integration time step, $\delta t$, between the SChPEq model and the exact ACKS2 solution.
    The two lower panels (c and d) also show the size of the charge and XL-DOFs. Mean absolute errors are computed from 1 ps of an equilibrated MD trajectory of a water molecule. All fits are against a log-base 10 abscissa.}
\centering
\includegraphics[width=1.0\textwidth]{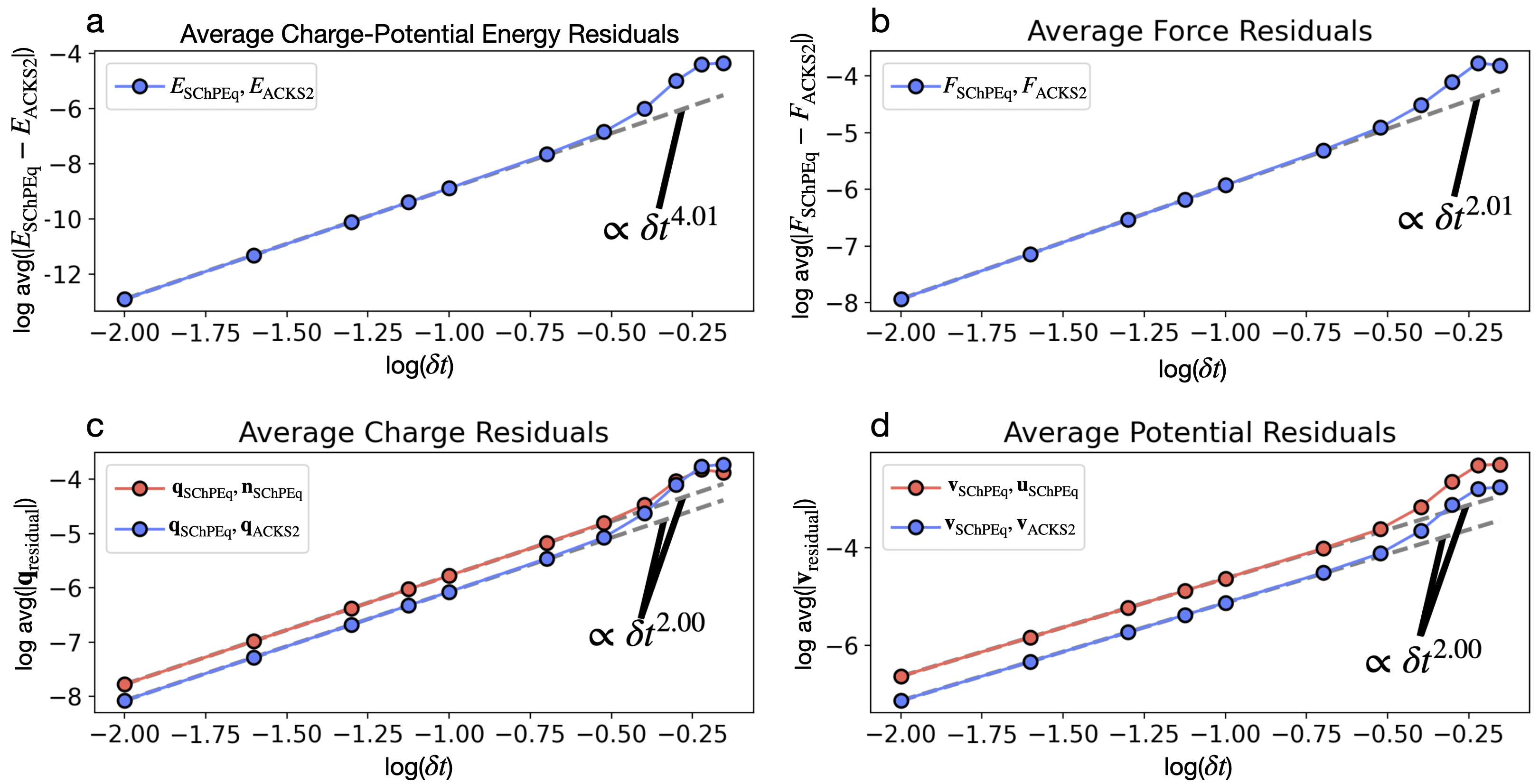}
\label{fig:scaling}
\end{figure}

Figure \ref{fig:scaling} demonstrates the nearly ideal empirical agreements with these expected scalings as a function of the time step for a single water molecule over a 1 ps MD trajectory in the NVE ensemble. A more detailed discussion on the scaling of the errors in shadow MD is given for the original formulation of the XL formalism\cite{Aniklasson21b}. As will be demonstrated in subsequent sections, these scaling relationships exhibit extensibility to larger systems. 
Figure \ref{fig:scaling} shows not only show that the error can be rigorously controlled by tuning the size of the time step, but also that the magnitudes of the errors are exceedingly small for reasonable selections of the integration time step. For example, the error in the charge-dependent potential Fig. \ref{fig:scaling}a is around 10$^{-6}$ eV/Atom for an integration time step around 0.5 fs.   

\begin{figure}[hbt!]
\caption{Dynamical properties of the SChPEq for a 25 ps microcanonical (NVE) simulation of a system of 10 waters at approximately 300 K using a time step of 0.25 fs. (a) The charge dependent portion of the potential energy for the SChPEq and ACKS2 system, exhibiting good agreement with one another. (b) total (kinetic + potential) energy (red) and SChPEq energy (black) across the 25 ps simulation, (c) comparison between the charge $q$ and propagated XL-DOF $n$ for an oxygen and hydrogen atom in the MD trajectory, (d) comparison between the potential $v$ and propagated XL-DOF $u$ for an oxygen and hydrogen atom in the MD trajectory.}
\centering
\includegraphics[width=1.0\textwidth]{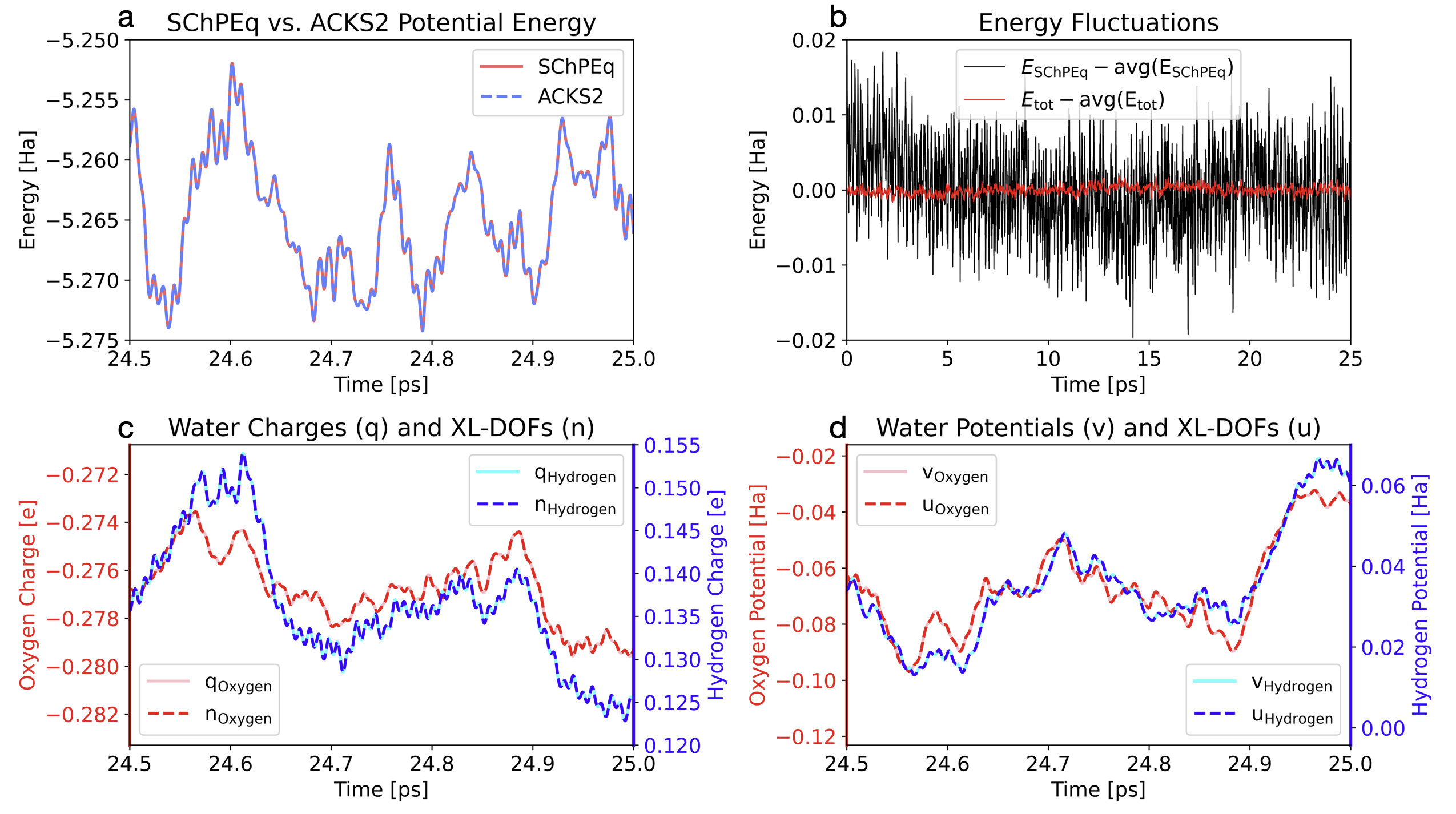}
\label{fig:overtime}
\end{figure}

The above analysis provides excellent agreement between the theoretical and realized scaling of errors with $\delta t$ for a small model system of a single water molecule.
Of importance, however, is also the realization of stable dynamics and the accuracy of the XL-DOFs in a more complex system over a longer period of time.
To this end, we investigate the dynamical properties of a water droplet containing 10 water molecules thermalized to approximately 300 K over 25 ps with a time step of 0.25 fs.
Figure \ref{fig:overtime}a demonstrates the close agreement between the SChPEq and direct ACKS2 potential energies. The mean absolute error (MAE), of the difference between the ACKS2 and SChPEq energy, $|E_{\rm ACKS2}-E_{\rm SChPEq}|$,  coincide with those presented in Figure \ref{fig:scaling} (MAE$=2.3 \cdot 10^{-9}$ Ha over the entire trajectory), which demonstrates promising extensibility of the theoretical scaling relationships in Figure \ref{fig:scaling} to larger systems. Figure \ref{fig:overtime}b shows the total energy (kinetic + potential) fluctuations and fluctuations in the charge dependent part of the SChPEq potential. The fluctuations, computed as the standard deviation $\sigma$, in the total energy are nearly exactly an order of magnitude lower than those in the SChPEq potential energy ($\sigma(E_{\rm tot})=5.39 \cdot 10^{-4}$ Ha vs. $\sigma(E_{\rm SChPEq})=5.38 \cdot 10^{-3}$ Ha), and no significant energy drift is observed in the total energy over the 25 ps, 100,000 time step trajectory.
The time series associated with the total energy exhibits no drift, and is stationary with a test statistic of -15.6 (1\% critical value of -3.43), and p-value of $< 10^{-16}$ for the Augmented Dickey-Fuller (ADF) test.
The ADF test offers more rigorous support for the time-reversibility of dynamics using the shadow potential than the typical visual inspection for the lack of drift in energy. This is accomplished by ensuring constant variance in energy fluctuations which preclude, for example, non-conservative transitions between nearby local minima of the potential energy surface of the system.

As seen in Figure \ref{fig:overtime}c, d, following the charges and potential fluctuations of a single oxygen and a hydrogen atom, the residuals between the XL-DOFs and their self-consistently solved counterparts are negligible, suggesting that a single (or very intermittent) solution of the initial preconditioner is sufficient to propagate the XL-DOFs in a stable manner. It is the benefit associated with this low rank update, together with the fact that the quasi-diagonal system of equations in the SChPEq formalism admits a solution that can leverage the Woodbury formula that leads to the reduced cost of SChPEq dynamics over the base ACKS2 model \cite{MR1927606}. These benefits are particularly important in large systems where the iterative solution of the ACKS2 system of equations becomes expensive.

\subsection{3.2 Fragment Charges in Flexible Charge Models}

\begin{figure}[h]
    \caption{Charge and potential fluctuations along a short MD trajectory with a .25 fs time step a) Schematic representation of the trajectory taken by the $H_2$-hydrogen atom. Near the trajectory endpoints, the system is composed of an H$^+$ and OH$^{-1}$, and near the center the system is a neutral H$_2$O molecule. b) The atomic partial charges from SChPEq, ACKS2, and DFT. c) The potential fluctuations from the SChPEq and ACKS2 models.}
\centering
\includegraphics[width=1.0\textwidth]{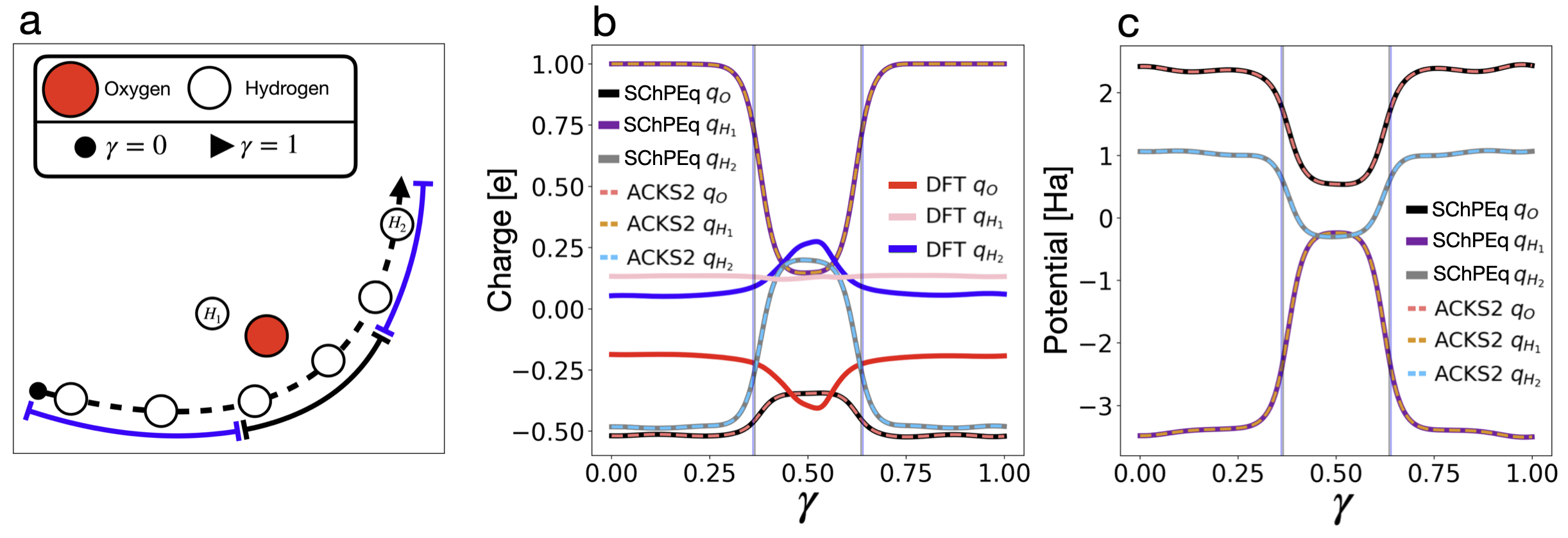}
\label{fig:flyby}
\end{figure}

A key shortcoming of both regular QEq models and of DFT (without spin) is the prediction of unphysical fractional partial charges on spatially isolated molecular fragments. The $\bf{X}$ term in the ACKS2 model is the key to remediating this issue. In this section we demonstrate that the SChPEq framework maintains this favorable characteristic of the ACKS2 model. The action of the potential fluctuation terms, $\bf{v}$, can be seen as Lagrange multipliers that, in the limit of an increasingly spatially isolated atom, serve to constrain the equilibrated charge to the reference charge specified in $\bf q_{\rm 0}$. Figure \ref{fig:flyby}a schematically represents a brief MD trajectory of a proton following a curved trajectory around a hydroxyl ion. A time step of 0.25 fs was used for all subsequent MD simulations presented in this study.  \\

Three qualitatively different phases are observed along the trajectory.
In the first phase we have separated $\rm H^+$ and $\rm OH^-$ ions, whereby we see that the ACKS2 and SChPEq models, are in agreement, capturing the proper integer charge of the molecular fragments (Figure \ref{fig:flyby}b).
This is in contrast to the spin-paired DFT results which display fractional partial charges.
Subsequently, we see the brief formation of an $\rm H_2O$ molecule, at which point the charges from the ACKS2 and SChPEq model exhibit near agreement with DFT calculations.
Finally, in the last phase, the dissociation of the water molecule occurs, and the ACKS2 and SChPEq models again correctly capture the integer charge on the $\rm H^+$ and $\rm OH^-$ fragments. Figure \ref{fig:flyby}c shows the on-site potential fluctuations, ${\bf v}$, associated with the $\bf X$ portion of the system of linear equations in Equation \ref{eq:3}. The large potential fluctuation on the $\rm H^+$ ion in its isolated form is the mechanism by which, in the ACKS2 and SChPEq models, isolated charge fragments are penalized for deviations from their references charges, $\bf q_{\rm 0}$. This can be seen most clearly by considering the structure of the $\bf X$ term from Eq.\ (\ref{eq:1}) when no other atoms are nearby. In all cases, we see nearly identical values between the ACKS2 and SChPEq models, consistent with Figure \ref{fig:scaling}. This highlights why integer charge fragments are a problem not only for a straightforward QEq approach, but also for more expensive electronic structure frameworks like DFT, which lack a penalization mechanism to mitigate the unphysical electron delocalization across spatially separated fragments. Given this improved physical fidelity, the stability and the reduced computational cost of the SChPEq framework over ACKS2 for dynamical simulations makes it a promising approach for MD simulations of reactive or fragmented systems, particularly when low-density intermediary phases are to be expected.

\subsection{3.3 Molecular Polarizabilities in the SChPEq Framework}

Another key problem associated with the regular formulation of the QEq framework is its scaling of macroscopic polarizabilities with system size \cite{verstraelen2014direct}. This is of particular importance for applications concerning photonics, organic semiconductors, and photoactive molecules used for biomedical signalling applications in drug delivery or medical imaging \cite{hosseinzadegan2018application,poszwa2022geometry,booth2022modelling,steele2010experimental}. 
For these applications, it is often the case that field-matter interactions dictate the efficiency of the corresponding application. 
This means that having computational frameworks which offer accurately predictable polarizabilities with system size (e.g. with particle radius in the case of drug delivery, or slab thickness in the case of semiconductors) becomes increasingly important to ensure correspondence between computational and experimental results. Here, we demonstrate that while the ACKS2 model mitigates the incorrect cubic scaling inherent in the standard QEq approach, in its current implementation, the ACKS2 model does not fully resolve the issue. \\

\begin{figure}[ht!]
    \caption{a) Molecular polarizabilities vs alkane chain length for the ACKS2 and QEq models, with the smallest and largest alkane chains pictured in the inset. b) Log-log plot of the molecular polarizabilities vs alkane chain length, demonstrating improvements of the ACKS2 model over the nearly cubic scaling in the QEq model. Alkane chains presented here follow the standard chemical formula C$_{\rm N_{\rm Carbon}}$H$_{\rm 2N_{\rm Carbon+2}}$. }
\centering
\includegraphics[width=1.0\textwidth]{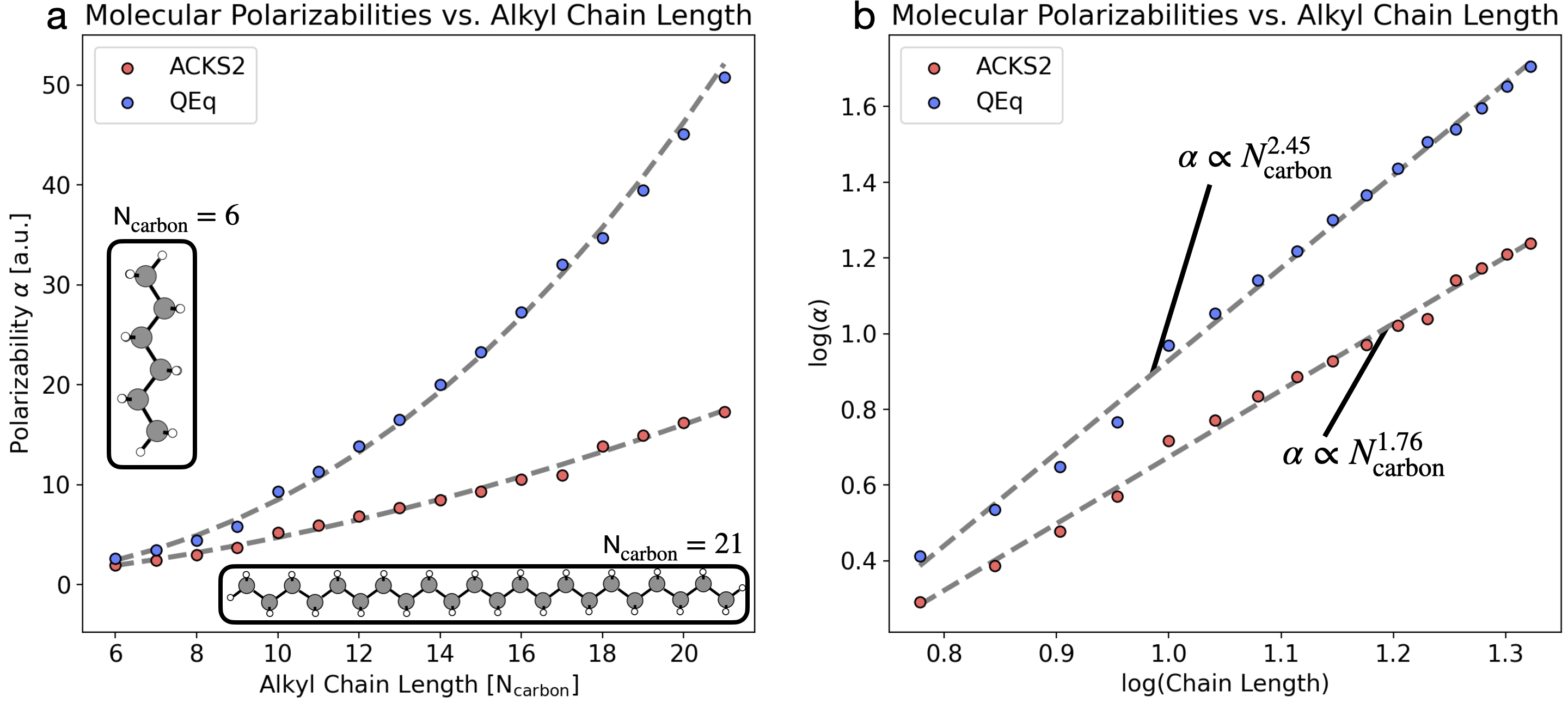}
\label{fig:polar}
\end{figure}

The regular QEq and ACKS2 scaling of molecular polarizabilities for alkanes with chain lengths of between 6 and 21 carbons are pictured in Figure \ref{fig:polar}a and in a log-log plot in panel b. The polarizabilities were calculated with a mid-point finite difference of the dipole moment with respect to a change in the strength of the electric field, where the electric field was applied along the direction of the alkane chain. We find that the regular QEq model does, indeed, exhibit nearly cubic scaling in the polarizability ($\alpha \propto N_{\rm carbon}^{2.45}$). Only the comparison between the ACSK2 model and QEq model is used (as opposed to the SChPEq model) due to the fact that the XL-DOFs in the SChPEq model are not well defined outside of the context of MD simulations. However, we assume that the SChPEq model will capture the same properties as the ACKS2 model during a MD simulation.
Figure \ref{fig:polar} shows that the ACKS2 model mitigates the nearly cubic scaling of the polarizability in the QEq model, although it still results in a superlinear scaling of the dipole polarizability with system size, where $\alpha \propto N_{\rm carbon}^{1.76}$. This issue may partially arise because the current implementations of the ACKS2 model contain only a monopole expansion of the charge fluctuations. While outside the scope of the present work, higher order terms (and in particular those resulting in the inclusion of on-site atomic dipoles) may aid in mitigating this superlinear scaling as the additional degrees of freedom can serve to screen the long range Coulomb interaction that occurs in the monopole-only implementation. It also appears to be the case that the inclusion of additional degrees of freedom through geometry-aware machine learning methods can mitigate this issue as well \cite{Shao_2022}.
In the next section we examine the ability of the SChPEq model to describe vibrational properties, including the infrared (IR) spectra.

\subsection{3.4 Dynamic Properties of the Charges}


By introducing partial charges and potential fluctuations as additional extended dynamical variables in our shadow MD we create a fictitious dynamical system beyond the motion of the nuclei on the regular Born-Oppenheimer potential energy surface. Even if the extended dynamical variables closely follow the exact fully optimized charges and potential fluctuations of the ACKS2 reference model, their dynamical properties may still deviate. In particular, artificial oscillations could potentially lead to resonances in the dipole moments that may alter the behaviour of the MD simulations. By calculating the IR spectrum from the dipole autocorrelation function sampled from MD simulations we can investigate if the IR spectra from the SChPEq model deviates from the ACKS2 reference model. This could reveal if the extended charge and potential fluctuation variables in the extended Lagrangian shadow MD lead to any significant problems.

\begin{figure}[ht!]
    \caption{Geometric and vibrational properties for the 33 water system computed along a 25 ps trajectory. a) OH bond length distribution, b) HOH bond angle distribution, c) IR spectra for the 33 water droplet using the 'exact' ACKS2 model in comparison the IR spectra for the SChPEq model using the XL-DOF charges, and d) experimental water droplet spectra, reproduced with permission from ref. 102.}
\centering
\includegraphics[width=1.0\textwidth]{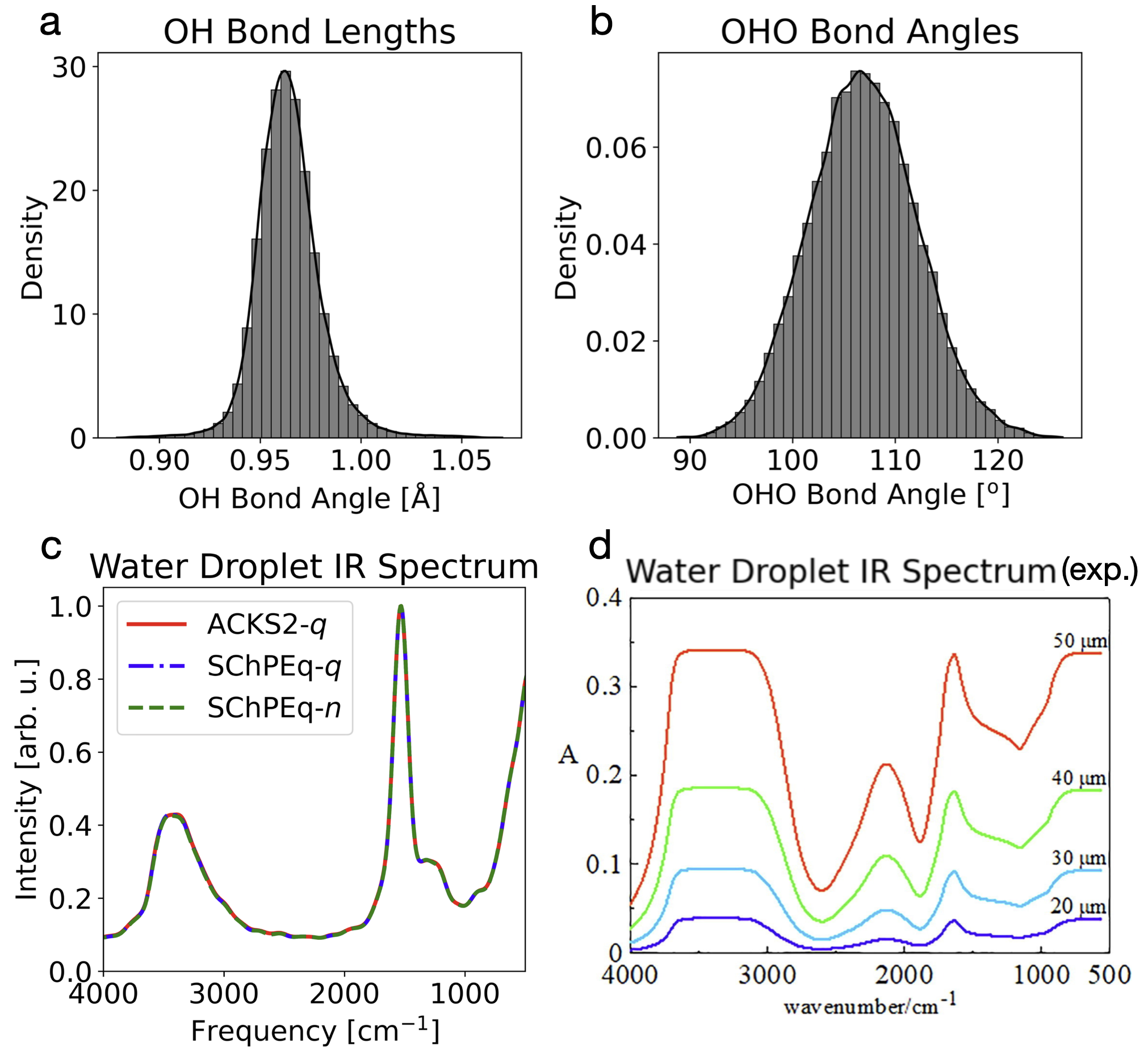}
\label{fig:spec}
\end{figure}

Figure \ref{fig:spec} demonstrates data sampled from an MD simulation of 33 water molecules at approximately 300 K for 25 ps, including OH bond lengths, OHO bond angles, and the IR spectrum computed from the dipole autocorrelation functions using the 'exact' equilibrated ACKS2 charges, the ground-state charges from the SChPEq method, and the extended dynamical charges, ${\bf n}(t)$, in the SChPEq framework.

Figure \ref{fig:spec}a, b shows sampled data of the of the OH bond length distribution and the HOH bond angle distribution.  The sampled data overlap reasonably well with experimental observations of ideal OH bond lengths and OHO bond angles of 0.96 Å and 104.5$^{\rm o}$ respectively.
However, more important for gaining an understanding of the extended Lagrangian SChPEq model are the IR spectra diplayed in Figure \ref{fig:spec}c. This panel displays the IR spectra calculated with the dynamical SChPEq charges, ${\bf n}(t)$, and the optimized ground state charges, ${\bf q}$ of the SChPEq model, in comparison to the ACKS2 reference model. All three spectra are virtually identical to one other. No artificial peaks appear because of the fictitious dynamics of 
${\bf n}(t)$ and ${\bf u}(t)$ in the shadow dynamics of the SChPEq model. The dynamical properties of the charges in the shadow MD therefore accurately represents the ACKS2 reference model. 

Even if the main purpose of our analysis is to demonstrate consistency between the shadow MD and the ACKS2 reference model,
it is still interesting to compare the spectra seen in Figure \ref{fig:spec}c with the experimental spectra for water droplets shown in Figure \ref{fig:spec}d \cite{droplet}.
Our results align nicely with those presented by Xu \textit{et. al} \cite{droplet}. At first glance, the plateau in the spectra centered around 3400 cm$^{-1}$ appears to be small and the broad peak near 2200 cm$^{-1}$ appears to be missing. However upon further inspection of Figure \ref{fig:spec}d, and other figures from the manuscript of Xu \textit{et. al}, one can see a decrease in the width of the peak near 3400 cm$^{-1}$ and a relative diminishing of the peak at 2200 cm$^{-1}$ as one decreases the diameter of the water droplet. The smallest water droplet in this experimental investigation was 1 $\mu$m, relative to approximately $10^{-3 }\ \mu$m in our MD simulation, allowing only such an extrapolatory comparison. Nevertheless, this trend supports the physical accuracy of the IR spectra from the charge-potential fluctuation model. These results also illustrate how the charge-potential equilibration model can be extended and used to analyze properties beyond those used for the parametrization of the model.



\section{Conclusion}


We have introduced the SChPEq framework for extended Lagrangian shadow MD simulations based on the second-order atom-condensed Kohn-Sham charge-potential equilibration (ACKS2) method. This shadow MD approach enables stable and computationally efficient simulations while maintaining close fidelity to the ACKS2 reference model. The shadow potential formulation allows for an exact direct solution, eliminating the need for costly iterative solvers, which, if only approximately converged, can introduce energy errors, non-conservative forces, and long-term instabilities in MD simulations.

Furthermore, we have demonstrated that differences in energies, forces, and partial charges between the ACKS2 and SChPEq models can be easily tuned by adjusting the integration time step, $\delta t$, used in the MD simulations. We have also shown that the ACKS2 and corresponding SChPEq models help address some limitations of DFT and standard QEq models, particularly regarding molecular polarizabilities and partial charges on molecular fragments.

Finally, we have demonstrated that the SChPEq MD simulations accurately captures dynamic properties of the charges in close agreement with the ACKS2 reference model, as was seen in the IR spectra calculated from the dipole auto-correlation function.  


\section{Computational Methods}

All DFT calculations used to obtain reference charges for the ACKS2 and SChPEq fitting procedures were carried out using the GPAW package\cite{mortensen2024gpaw}, and employed the Perdew Burke and Ernzerhof (PBE) functional\cite{perdew1996generalized}. A double $\zeta$-valence polarized basis set was used, and all population analysis for atomic partial charges was carried out using the Hirshfeld scheme \cite{larsen2009localized}. No isolated fragments were used in the the generation of reference data to avoid the erroneous partial charges obtained from DFT in such systems. Self-consistent field calculations were carried out to within $0.5$ meV/valence electron and $10^{-4}$ $e$/valence electron for the energy and density, respectively. \\
Simulations using the ACKS2 and SChPEq system require a charge neutral portion of the potential to obtain forces which do not directly arise from the atom-condensed partial charges. In other words, the total energy function for all systems used in the present study are of the form $E^{\rm tot}({\bf R,q,v})= V_{\rm S}({\bf R}) + E({\bf q,v})$, where $V_{\rm S}({\bf R})$ represents the short-range charge-neutral potential energy, and $E({\bf q,v})$ represents the charge-potential energy function in Eq.\ (\ref{ACKS2Energy}). For all simulations, $V_{\rm S}({\bf R})$ is approximated using the charge-independent portion of the potential energy in the GFN0-xTB formalism \cite{pracht2019robust}.

\begin{acknowledgement}
This work is supported by the U.S. Department of Energy Office of Basic Energy Sciences (FWP LANLE8AN), the LANL LDRD program, and by the U.S. Department of Energy through the Los Alamos National Laboratory (LANL). LANL is operated by Triad National Security, LLC, for the National Nuclear Security Administration of the U.S. Department of Energy Contract No. 892333218NCA000001. Additionally, this research used resources provided by the Los Alamos National Laboratory Institutional Computing Program (the Chicoma cluster), which is supported by the U.S. Department of Energy National Nuclear Security Administration under Contract No. 89233218CNA000001. Finally, we thank Chen-Han Li for his assistance in obtaining the computational IR spectra. LA-UR-25-21258.
\end{acknowledgement}

\newpage
\bibliography{References_RS,References_AMN}

\end{document}